\documentclass[10pt,a4paper,twocolumn]{article}
\usepackage[utf8]{inputenc}
\usepackage[T1]{fontenc}
\usepackage{lmodern}
\usepackage[a4paper,margin=2.0cm,columnsep=0.65cm]{geometry}
\usepackage{amsmath,amssymb,amsfonts}
\usepackage{graphicx}
\usepackage{textcomp}
\usepackage{xcolor}
\usepackage{float}
\usepackage{caption}
\usepackage{authblk}
\usepackage[colorlinks=true,linkcolor=blue,citecolor=blue,urlcolor=blue]{hyperref}
\IfFileExists{xurl.sty}{\usepackage{xurl}}{}
\usepackage{booktabs}
\usepackage{tikz}
\usetikzlibrary{arrows.meta}
\IfFileExists{siunitx.sty}{%
  \usepackage{siunitx}
  \sisetup{separate-uncertainty=true}
}{}
\providecommand{\sisetup}[1]{}
\providecommand{\SI}[2]{#1\,#2}
\providecommand{\si}[1]{#1}
\providecommand{\byte}{B}
\providecommand{\bit}{bit}
\providecommand{\mega}{M}
\providecommand{\kilo}{k}
\providecommand{\kibi}{Ki}
\providecommand{\giga}{G}
\providecommand{\hertz}{Hz}
\providecommand{\milli}{m}
\providecommand{\micro}{\ensuremath{\mu}}
\providecommand{\second}{s}

\usepackage[backend=biber, style=ieee, isbn=false, sortcites, maxbibnames=6, minbibnames=1, safeinputenc]{biblatex}
\DeclareSourcemap{
  \maps[datatype=bibtex]{\map{\step[fieldsource=abstract,null]\step[fieldset=abstract,null]}}
}

\AtBeginBibliography{\small}

\title{\textbf{Entropy Bootstrapping for Wireless Embedded Systems}}
\author[1]{Francisco Javier Blanco-Romero\thanks{Corresponding author: frblanco@pa.uc3m.es}}
\author[1]{Florina Almenares Mendoza}
\author[1]{Daniel D\'iaz-S\'anchez}
\author[2]{Andr\'es Mar\'in-L\'opez}
\affil[1]{Department of Telematic Engineering, Universidad Carlos III de Madrid, Spain}
\affil[2]{Department of Telematic Systems Engineering, Polytechnic University of Madrid, Spain}

\begin{document}

\makeatletter
\twocolumn[
  \begin{@twocolumnfalse}
    \maketitle
    \begin{center}
      \small\textbf{Abstract}
    \end{center}
    \begin{quote}
      \small
Cryptographic protocols require unpredictable randomness at first boot, yet wireless sensors often wake with uninitialized hardware or noise sources running in unverified operating states. In platforms like the ESP32, statistical testing cannot catch this cold-start failure because the internal random number generator continues returning statistically plausible bytes even when analog radio noise is disabled, producing pure pseudorandomness by design. To prevent nodes from silently anchoring security in a single unverified source, we introduce a defense-in-depth boot architecture governed by explicit source-state admission. The system combines three orthogonal roots of unpredictability: startup noise from uninitialized SRAM cells, local wireless RNG output gathered during a bounded window defined by an external packet burst, and a signed post-quantum asymmetric capsule delivering fresh entropy from a trusted peer without requiring local random generation. Rather than blending unverified numbers, the admission policy gates entropy credit on the verified operating mode of the hardware, admitting wireless RNG output only while the radio frequency circuits are active. Empirical trials confirm that incoming network traffic successfully bounds repeatable, uncorrelated hardware sampling windows, while remote capsule verification completes with minimal processing overhead. By coupling physical operating state to cryptographic accounting, commodity microcontrollers can bootstrap reliable seed entropy across complementary roots without relying on additional dedicated security hardware.
    \end{quote}
    \vspace{4ex}
  \end{@twocolumnfalse}
]
\makeatother


\section{Introduction}

Secure embedded protocols require unpredictable randomness before their first network exchange, yet hardware entropy sources may still be uninitialized or in unverified operating states at cold start. A device wakes, initializes drivers, joins a wireless network, and asks a random number generator (RNG) for values that must already be unpredictable. If these first values are weak, later cryptography inherits the weakness.

Several deployed systems have suffered serious randomness failures. The Debian OpenSSL bug restricted generated keys to a small, enumerable set~\cite{yilek2009private}. Later Internet scans traced shared and factorable RSA keys to headless devices generating keys at first boot, while smart cards repeated primes~\cite{heninger2012mining,mowery2013welcome,bernstein2013factoring}. Dual~EC and Juniper demonstrate the related risk of deliberate backdoors~\cite{checkoway2014practical,checkoway2016systematic}. In each case, plausible output concealed a failure in the source.

This failure mode is relevant to embedded devices. Thousands of units may run identical firmware on identical silicon, enter the same reset vector, and perform the same short network task. The RNG API can keep returning plausible bytes even when the physical source behind it has changed state, a failure mode seen in firmware studies~\cite{tillmanns2020firmware}. On ESP32 devices, for example, the RNG register still returns bytes when documented analog entropy inputs are inactive. In that state, the output is pseudorandom by specification~\cite{espressif2026esp32trm}. Section~\ref{sec:esp32_trng_characterization} shows that this rejected state passes the same statistical screens as states with radio frequency (RF) circuitry active. Output tests can support a source model, but they cannot replace it.

We address this problem by combining inputs with distinct failure modes. SRAM contributes startup state captured on the board. Radio extraction samples the wireless RNG during a public receive window, while an entropy capsule supplies remote material protected by provisioned asymmetric keys. The radio and capsule paths require a trusted node in the local Internet of Things (IoT) network, such as a gateway or dedicated entropy node. The seed remains unpredictable when at least one credited root remains unknown to the adversary.

We define an admission policy that assigns credit per boot, then derive the seed from every available input without adding their claims. We implement the radio burst and capsule paths as open Zephyr artifacts. Full experiments use one ESP32-DevKitC~V4, with selected source state, burst, and SRAM controls repeated on one ESP32-S3 DevKitC-1.

\section{Related Work}
\label{sec:related}

Weak key scans and direct boot time measurements establish the cold start problem for headless devices~\cite{heninger2012mining,mowery2013welcome}. Robust pool constructions~\cite{barak2005model} and IoT pseudorandom number generator (PRNG) guidelines~\cite{kietzmann2021guideline} address conditioning and recovery, while entropy source standards separate tests from credit. NIST Special Publication (SP)~800-90B requires a model of the noise source, its expected entropy, and its health tests~\cite{sonmez2016recommendation}. AIS~20/31 makes source assumptions explicit through functionality classes~\cite{peter2024ais2031}. Saarinen likewise shows why statistical screens cannot establish cryptographic unpredictability~\cite{saarinen2022sp}. The common requirement is a known operating state, not merely plausible output.

Entropy as a Service moves the source off the device and delivers randomness to systems that cannot gather enough locally~\cite{vassilev2016entropy}. Our previous prototype delivered entropy from a quantum random number generator (QRNG) to ESP32 devices over the Constrained Application Protocol (CoAP) and Datagram Transport Layer Security (DTLS)~\cite{blanco2026post}. With a provisioned client public key, such a service can cover cold start as well as later refills, although the server knows the delivered bits. This remote trust model differs from RFC~8937, which mixes long term private key material into protocol randomness on the device~\cite{rfc8937}.

At the physical layer, key generation schemes derive shared keys from channel measurements made by two nearby radios that an outside observer cannot reproduce exactly~\cite{mathur2008radio,mukherjee2014principles,jiao2019physical}. Induced randomness and reconfigurable intelligent surface (RIS) variants add controlled perturbations when the channel is too static~\cite{aldaghri2019physical,yang2022risinduced,li2022risconstructive}.

Wireless true random number generator (TRNG) designs instead embed the source in the communication system~\cite{suzuki2018wirelesstrng}. SRAM power up state provides another hardware source and has been used for both device identification and random number generation~\cite{holcomb2008power,guajardo2007fpga,herder2014physical}.

\section{Problem and Threat Model}
\label{sec:model_problem}

\subsection{Boot Entropy Starvation}

At first boot, a node may need random values before it has evidence that any local source is ready. DTLS~1.3 consumes randomness in fresh protocol values and ephemeral key exchange~\cite{rfc8446,rfc9147}. KEM protocols also require randomness because encapsulation is randomized, although decapsulation is deterministic~\cite{rfc9180,nist2024fips203}. A provisioned symmetric key shifts part of the problem to provisioning, storage, and rotation, while protocols still need fresh local values to avoid deterministic reuse.

The first seed must therefore depend on at least one input that the adversary cannot determine. We call such an input a root of unpredictability. It may provide secrecy because the value is hidden from a remote adversary, freshness because it changes from boot to boot, or both. Network messages can coordinate the protocol and record what happened, but cannot by themselves create a secret seed.

\subsection{Adversary}

We assume a strong local adversary. It observes all radio traffic, knows the firmware, extraction functions, access point configuration, schedules, payload patterns, and nonces, and can replay, delay, drop, inject, interfere, and stand near the node. It does not have code execution and cannot read internal memory or registers. It may try to degrade local entropy sources, for example by jamming, saturating the receiver front end, or injecting periodic signals into a physical noise source~\cite{markettos2009frequency}.

The adversary wins if it can predict the node's first seed with non-negligible advantage. For a root of unpredictability $R$ collected during public events $S$, the requirement is
\[
H_\infty(R \mid S, O_A) \geq k,
\]
where $O_A$ is the adversary's observation and influence of the RF and network environment, and $k$ is empirical and specific to the platform. Because we do not assume a value of $k$, we evaluate each root separately. Rejecting one root leaves the seed dependent on another credited root that remains unpredictable to the adversary.

\section{Architecture}
\label{sec:architecture}

At boot, the combiner derives a seed from every available input, while entropy credit records which inputs meet their admission conditions. Inputs with zero credit can still enter the combiner, and a device may request a capsule on every boot even when local material is available.

\subsection{SRAM Startup State}
\label{sec:puf_root}

Uninitialized SRAM can contain both stable cell preferences and cells that vary between power cycles~\cite{holcomb2008power,guajardo2007fpga}. Prior work uses these properties for device fingerprints and startup randomness when the capture path and power conditions are controlled. Our combiner simply hashes the observed bytes; it does not reconstruct a stable SRAM key. A deployment that needs stable identity can add a helper-data or fuzzy-extractor construction~\cite{dodis2004fuzzy,herder2014physical}. Entropy credit, however, requires a capture after complete power removal and before initialization of the region. Our measurements do not meet those conditions, so SRAM enters only as auxiliary material.

\subsection{Radio Burst Extraction}
\label{sec:rf_root}

The second root uses controlled radio traffic to open a collection window. After receiving a public burst descriptor and packet train, the sensor records WDEV output and local packet arrival timing and derives its contribution internally. Security does not come from the burst itself. An observer who knows the complete stimulus must still predict the measurements inside the client.

On the ESP32, Wi-Fi association places WDEV in its documented RF-active entropy state because the radio enables the high speed ADC. Thermal-noise bit streams from that ADC are mixed into the RNG state~\cite{espressif2026esp32trm}. The burst does not create that state; it provides a bounded, repeatable window for collecting the device's local response. Every predictability trial therefore uses the same public burst. Section~\ref{sec:rf_model} defines this measurement, and Section~\ref{sec:aeb_source_separation} reports the results.

\subsection{Asymmetric Entropy Capsules}
\label{sec:capsule_root}

\begin{figure}[t]
\centering
\scriptsize
\begin{tikzpicture}[font=\scriptsize,>=Latex]
\node[align=center] (client) at (0,0) {\textbf{Client}\\before DTLS};
\node[align=center] (trusted) at (6.3,0) {\textbf{Trusted node}\\approved entropy};
\draw (client.south) -- ++(0,-5.0);
\draw (trusted.south) -- ++(0,-5.0);
\draw[->] (0,-1.0) -- node[above,align=center,text width=4.9cm,fill=white,inner sep=1pt]
  {\textbf{\texttt{BOOT\_HELLO}}\\$id_D, ctr_D, profile, H(local)$} (6.3,-1.0);
\node[align=center,text width=4.9cm,fill=white,inner sep=1pt] at (3.15,-2.1)
  {\textbf{trusted randomized step}\\check client registration; encapsulate to $pk_D$; sign};
\draw[<-] (0,-3.2) -- node[above,align=center,text width=4.9cm,fill=white,inner sep=1pt]
  {\textbf{\texttt{CAPSULE}}\\$id_D, ctr_D, seq, t_G, c, H(hello), \sigma_G$} (6.3,-3.2);
\node[align=center,text width=5.0cm,fill=white,inner sep=1pt] at (3.15,-4.35)
  {\textbf{client deterministic step}\\verify signature and transcript; decapsulate; derive key};
\end{tikzpicture}
\caption{Asymmetric entropy capsule exchange. The randomized cryptographic work runs on the trusted node. The client verifies, decapsulates, and derives the first pool input before DTLS.}
\label{fig:capsule_protocol}
\end{figure}
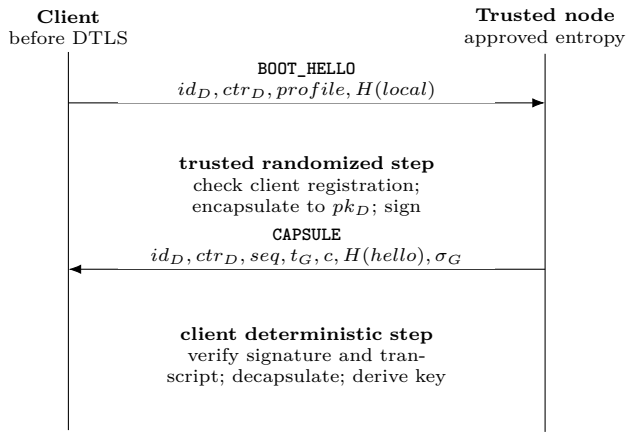

The third root uses provisioned asymmetric key material so capsule processing does not require the client to generate randomness. Before deployment, the device receives a decapsulation private key, or a protected seed from which that key is derived, plus a trust anchor for the trusted node. The trusted node holds or obtains the matching client public key. The capsule exchange in Figure~\ref{fig:capsule_protocol} runs before DTLS over the User Datagram Protocol (UDP) or CoAP over UDP. The client performs deterministic verification, decapsulation, and hashing before accepting the seed~\cite{nist2024fips203,nist2024fips204}. The prototype joins Wi-Fi and obtains an address before requesting a capsule. Any randomness required for network attachment must be supplied separately.

The capsule provides an independent remote root that can be combined with local inputs or used when local sources are unavailable. Because the trusted node knows the delivered material, it protects against outside observers but not against a compromised trusted node. When this is the only root, ML-KEM-512 limits confidentiality to roughly 128 bits, so the isolated benchmark injects 32 bytes and configures at most 128 bits of credit. A production algorithm profile should match the target security level and combine ML-KEM with a classical mechanism such as X25519.\footnote{The prototype isolates ML-KEM and ML-DSA to measure the post-quantum path. Deployed protocols should keep the classical component. Transport Layer Security (TLS)~1.3 already supports X25519, and RFC~7748 defines the curve operation~\cite{rfc8446,rfc7748,nist2024fips203}.}

At first boot, the client cannot safely use a locally generated nonce as proof of freshness because generating unpredictable values is the problem being solved. The prototype binds each capsule to the device identity, trusted node identity, algorithm profile, and current boot transcript. Preventing reuse of a valid capsule after a restart additionally requires a sequence number or counter stored in nonvolatile memory. The prototype does not implement that persistent replay state, so it authenticates capsule contents but does not establish capsule freshness across restarts.

\subsection{Admission Policy}
\label{sec:gating}

The device begins each boot with no credited entropy. The admission policy may credit WDEV output when RF is active, but assigns no credit when RF is disabled regardless of the byte distribution. A capsule provides separate credit only after signature, transcript, and freshness checks, and the device may request one even when local inputs are available. Cryptographic services become available when at least one admitted root supplies the target credit.

The PoC evaluates WDEV and capsule admission in separate experimental paths rather than in one automatic state machine. One path checks WDEV state, while the other authenticates the capsule transcript. Because the hardware resets did not remove board power and packet arrival timing has no conditional entropy estimate, both measurements enter only as uncredited auxiliary material. The PoC also records radio state, packet delivery, and timing but has no calibrated detector for jamming, injection, or receiver saturation.

\subsection{Seed Derivation and Credit Accounting}
\label{sec:combiner}

Let $P$ be the SRAM startup read, $R$ the radio burst response, $C$ the decapsulated capsule material, and $W$ direct chip RNG output under gating. An absent root is replaced by an empty string, and the transcript $T$ records its absence. We derive the seed with HKDF, the HMAC-based Extract-and-Expand Key Derivation Function.
\begin{align*}
\rho &= H(P \parallel R \parallel C \parallel W),\\
seed &= \mathsf{HKDF\mbox{-}SHA256}\big(\mathsf{ikm}=\rho,\ \mathsf{salt}=H(T),\\
&\qquad \mathsf{info}=\texttt{boot-entropy-combiner-v1}\big),
\end{align*}
where $T$ binds the public burst descriptors, capsule transcript, and root availability flags. Each root is encoded with an explicit length and a domain tag before concatenation~\cite{rfc5869}.

HKDF extracts a pseudorandom key from input material that may be nonuniform or partly influenced by an attacker, provided enough entropy remains after conditioning on public context~\cite{rfc5869}. Robust pools and PRNGs with input apply the same principle to continual mixing~\cite{barak2005model,dodis2013security}. Suppose at least one credited root remains hidden after conditioning on the public transcript and the other inputs. Under the usual random oracle heuristic, hashing the encoded roots before HKDF then preserves an unpredictable seed.

Credit follows the evidence for the inputs, not the length of the derived output. The seed therefore receives the largest supported credit of any individual root, capped at 256 bits and by the applicable cryptographic security bound. Adding the claims would require an independence argument for analog RNG noise, timing jitter, and SRAM mismatch on the same die. A production system should inject the derived seed into its normal random pool before applications consume randomness~\cite{aumasson2013blake2}.

\section{Radio Burst Measurement Method}
\label{sec:rf_model}

A radio burst measurement consists of a public burst plan $B_i$ and a local device response $R_i$. The node first sends a boot UDP \texttt{HELLO}. The collector replies with a \texttt{START} message and then zero or more \texttt{BURST} packets. The burst plan is
\[
\begin{aligned}
B_i = (&\mathsf{trial}_i,\mathsf{nonce}_i,\mathsf{burst\_count},\\
&\mathsf{interval},\mathsf{sample\_bytes}).
\end{aligned}
\]
The \texttt{HELLO}, nonce, schedule, and $\mathsf{trial}_i$ are public and contribute no entropy. The device instead records its local response. The main experiment holds the \texttt{START} nonce, burst count, spacing, and payload constant, so measured differences cannot be attributed to a changing public challenge.

The device response is
\[
R_i = W_i \parallel J_i,
\]
where $W_i$ is the sequence of chip RNG bytes sampled during the burst window and $J_i$ is the sequence of packet arrival deltas observed by the device. The firmware timestamps packets after UDP receive, so $J_i$ includes radio reception, driver, network stack, interrupt, and scheduling variation. It is not an isolated measurement of physical layer detection jitter.

For each trial, the firmware derives
\begin{align*}
\rho_i &= H(R_i), \qquad \tau_i = H(B_i),\\
seed_i &= \mathsf{HKDF\mbox{-}SHA256}\big(\mathsf{ikm}=\rho_i,\ \mathsf{salt}=\tau_i,\\
&\qquad \mathsf{info}=\texttt{radio-burst-entropy-v1}\big),
\end{align*}
which becomes the $R$ input of Section~\ref{sec:combiner}. The evaluation excludes the other roots and secure transport so that the measured path contains only the radio response~\cite{rfc5869}.

\section{Implementation}
\label{sec:implementation}

The artifacts are standalone Zephyr applications released as version v0.1.1.\footnote{\url{https://github.com/perlab-uc3m/wireless-boot-entropy-zephyr/releases/tag/v0.1.1}} Full experiments run on one ESP32-DevKitC~V4 with an ESP32-WROOM-32 module and revision~1.0 ESP32-D0WD silicon, which Zephyr warns is outside its supported set. One ESP32-S3 DevKitC-1 with revision~0.2 silicon repeats the limited source state comparison and both fixed burst experiments. Both boards use the same local \SI{2.4}{\giga\hertz} WPA2 network. The primary tests used a TP-Link Archer~AX73 router and a Dell OptiPlex~3000 collector.

On the primary ESP32, WDEV is the wireless device RNG path exposed by \texttt{WDEV\_RND\_REG} (also \texttt{RNG\_DATA\_REG}, \texttt{0x3FF75144}) and used by Zephyr's entropy driver~\cite{espressif2026esp32trm}. The radio tests bypass the custom BLAKE2s pool and read the stock entropy device on each board.

The radio burst path starts after the device under test (DUT) joins Wi-Fi and sends the public \texttt{HELLO}, which is distinct from the capsule \texttt{BOOT\_HELLO}. The collector script, \texttt{scripts/aeb\_collector.py}, answers with a binary \texttt{START} and sends the burst. After each valid packet, the DUT records the arrival delta and samples WDEV output. If packets are lost, it records the loss and completes the requested WDEV buffer after the packet loop. The timing stream contains only received packets.

Firmware logs hashes of the WDEV bytes, timing stream, combined response, and derived seed. Entropy test runs also upload unhashed WDEV bytes in binary UDP chunks. This keeps $W_i$, $J_i$, and $W_i \parallel J_i$ available for separate analysis.

To make the public stimulus reproducible, the collector records its nonce mode, burst count, interval, payload mode, and trial count in the run manifest alongside the WDEV trials, packet deltas, schedule residuals, joint responses, and stimulus hashes.

For source state characterization, the \path{esp32-rf-rng-state} harness captures bulk WDEV output under four controlled RF conditions. These are RF disabled with the Wi-Fi driver omitted, Wi-Fi idle after association and the Dynamic Host Configuration Protocol (DHCP), Wi-Fi scan with periodic scans, and a deterministic UDP burst matching the radio burst workload. The RF-disabled run provides the documented pseudorandom control.

For the startup measurements, the SRAM harness copies a dedicated \SI{4096}{\byte} \texttt{.noinit} region at Zephyr's earliest initialization level. One host command then builds and flashes the firmware, applies the pattern and reset schedule, saves captures, and calculates retention metrics.

The capsule prototype runs on ESP32 with local hardware refills disabled after initialization. The client sends an \SI{88}{\byte} \texttt{BOOT\_HELLO}, and the server returns a \SI{3252}{\byte} signed ML-KEM capsule bound to its hash. The logs separate network setup from server and client processing and record memory use and time to seed.

\section{Evaluation}
\label{sec:evaluation}

\subsection{ESP32 RNG State Evaluation}
\label{sec:esp32_trng_characterization}

On the primary ESP32, we captured \SI{256}{\mega\byte} streams with RF disabled and during associated idle, scan, and deterministic UDP burst conditions, using Linux \texttt{/dev/urandom} as a host control. We applied ENT~\cite{walker2008ent}, SP~800-90B~\cite{sonmez2016recommendation}, Borel normality, AIS31 P1/P2, GM/T~0005-2021, PractRand~\cite{dotyhumphrey2019practrand}, and TestU01 Rabbit~\cite{l2007testu01} through our \texttt{randlab} wrapper.\footnote{\url{https://github.com/perlab-uc3m/randlab}}

The main result in Table~\ref{tab:trng_results} is that output quality does not track the documented source state. All four ESP32 conditions have the same ENT score to six decimal places and comparable SP~800-90B non-IID estimates of 7.112--7.204 bits/B, including the RF-disabled state that the documentation identifies as pseudorandom. Rabbit likewise reports two or three flags in every ESP32 condition but none in the host control, with no relation to whether RF is active. Admission must therefore depend on the operating state reported by the driver rather than on a runtime statistical screen. The SP~800-90B values shown are $\min(H_\mathrm{original}, 8H_\mathrm{bitstring})$ across byte and bitstring views.

\begin{table*}[t]
\centering
\caption{ESP32 WDEV source state controls, \SI{256}{\mega\byte} per condition. RF-disabled output is documented pseudorandom but statistically similar to RF-active states. Rabbit counts are diagnostics.}
\label{tab:trng_results}
\scriptsize
\resizebox{\textwidth}{!}{%
\begin{tabular}{llcccl}
\toprule
Condition & Policy role & ENT entropy & SP~800-90B non-IID & Other screens & Rabbit flags \\
\midrule
RF disabled & Reject, documented pseudorandom & 7.999999 & 7.112 & ENT, Borel, AIS31, GM/T, PractRand pass & 3 \\
Wi-Fi idle & Eligible RF-active source state & 7.999999 & 7.204 & ENT, Borel, AIS31, GM/T, PractRand pass & 2 \\
Wi-Fi scan & Eligible RF-active source state & 7.999999 & 7.137 & ENT, Borel, AIS31, GM/T, PractRand pass & 3 \\
UDP burst & Target burst workload & 7.999999 & 7.198 & ENT, Borel, AIS31, GM/T, PractRand pass & 3 \\
\texttt{/dev/urandom} & Host control & 7.999999 & 7.044 & ENT, Borel, AIS31, GM/T, PractRand pass & 0 \\
\bottomrule
\end{tabular}
}
\end{table*}

The remaining batteries are still useful as regression checks, but they do not change that admission decision. On the ESP32-S3, the RF-disabled and associated Wi-Fi streams likewise produced nearly identical ENT values (7.999822 and 7.999835 bits/B) and throughputs (179,829 and 179,828~B/s). While not a substitute for a full battery across a larger population, this comparison confirms that output statistics alone remain blind to the underlying operating state.

\subsection{Radio Burst Response Evaluation}
\label{sec:aeb_source_separation}

The first radio burst experiment measures consecutive windows within one Wi-Fi association. Each trial used the same nonce, constant 64 byte payloads, a 64 packet burst at \SI{1000}{\micro\second} spacing, and 8192 unhashed WDEV bytes. The run produced 127 complete windows without a reset and received all 8128 expected packets; matching file and firmware hashes linked every saved WDEV and timing capture to its trial.

\begin{figure}[t]
\centering
\includegraphics[width=\columnwidth]{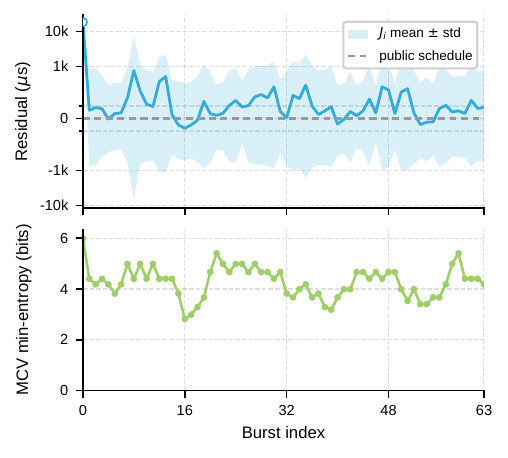}
\caption{Fixed public burst timing evidence. The programmed schedule residual is zero. The ESP32 curve shows the mean and one standard deviation of the node's timing residual over 127 trials. Position 0 includes the programmed \SI{20}{\milli\second} start delay. The most common value (MCV) curve is diagnostic and receives no entropy credit.}
\label{fig:aeb_burst_timing_entropy}
\end{figure}

Across 127 trials, the full 64 packet stimulus took \SI{89.48}{\milli\second} $\pm$ \SI{4.69}{\milli\second} from \texttt{START} to the last \texttt{BURST}. This short collection window includes the programmed \SI{20}{\milli\second} delay before the first packet, as shown in Figure~\ref{fig:aeb_burst_timing_entropy}, where the measured arrival residuals vary around the programmed zero-residual schedule.

The entropy estimate uses only WDEV bytes and is conditional on the RF-active source state model from Section~\ref{sec:gating}. Taking the lower of 7.348 bits/B from this run and 7.198 bits/B from the larger \SI{256}{\mega\byte} UDP burst benchmark gives approximately \SI{59}{\kilo\bit}, or \SI{7.20}{\kibi\byte}, across the \SI{8192}{\byte} window before extraction into a 256-bit seed. This extrapolation from the output stream assumes that the source state and distribution remain stationary on the same board and session; it is not a conditional min-entropy bound for an individual boot or an active attack and does not establish independence between windows. None of the 127 windows repeated, and consecutive windows had 49.952\% mean Hamming distance and 0.00138 mean Pearson byte correlation.

We then collected the first burst window after 129 RTS/EN resets and new Wi-Fi associations while USB power remained connected. A single association attempt failed; its retry yielded 128 complete windows in which every expected packet arrived and each saved capture matched its firmware hash. All WDEV hashes were distinct, and consecutive windows had 50.003\% mean Hamming distance and 0.000105 mean byte correlation. While the data reveal no obvious correlation between adjacent windows, they do not establish cryptographic independence and capture only reset behavior, as pulling EN low leaves the board supply energized.

The compiled \SI{1}{\mega\byte} WDEV stream from the same association passed all batteries, including TestU01 Rabbit. The unexplained flags in the larger \SI{256}{\mega\byte} runs still warrant conditioning raw WDEV bytes before use.

Repeating both experiments on the ESP32-S3 produced 127 complete association windows and 128 reset captures from 128 attempts, with 8188 of 8192 burst packets received and no repeated WDEV hash. Consecutive windows in the association and reset runs had mean Hamming distances of 49.996\% and 50.003\% and mean byte correlations of 0.000466 and 0.00153, respectively. The same variation appears on a second processor. The data do not establish independence or conditional entropy.

\subsection{Capsule Latency}
\label{sec:root_benchmarks}

\begin{table}[t]
\centering
\caption{Isolated capsule benchmark, ML-KEM-512 + ML-DSA-44, 30 RTS resets. Values are median [range].}
\label{tab:teb_boot_latency}
\scriptsize
\resizebox{\linewidth}{!}{%
\begin{tabular}{lc}
\toprule
Metric & Value \\
\midrule
Successful exchanges & 30 / 30 \\
Time to capsule seed & 7750 [3200, 13060] \si{\milli\second} \\
Wi-Fi association & 963 [962, 1063] \si{\milli\second} \\
DHCP / IPv4 & 5106 [1103, 9111] \si{\milli\second} \\
Fixed settle after address assignment & 1000 \si{\milli\second} \\
Capsule UDP exchange & 18 [10, 3000] \si{\milli\second} \\
Server capsule construction & 137.5 [90, 623] \si{\micro\second} \\
Client capsule processing & 36013.5 [35916, 36036] \si{\micro\second} \\
Verification / decapsulation / HKDF / injection & 12920 / 5195 / 600 / 963 \si{\micro\second} \\
Application traffic & 3340 \si{\byte} in 2 UDP datagrams \\
IPv4 packets at \SI{1500}{\byte} MTU & 1 request + 3 response fragments \\
Capsule input / configured credit & 32 \si{\byte} / 128 \si{\bit} \\
Peak heap / flash image / static DRAM & 48216 / 560420 / 184712 \si{\byte} \\
\bottomrule
\end{tabular}
}
\end{table}

The capsule benchmark isolates the remote path by disabling hardware refills and clearing initial local credit while retaining the local bytes as mixed material. A normal deployment may instead combine the capsule with every available input. Signature and transcript validation brought all 30 trials to the configured 128-bit accounting state. The volatile boot counter means that this benchmark does not establish freshness across restarts.

Network attachment set the median \SI{7750}{\milli\second} boot time. DHCP took \SI{5106}{\milli\second}, while Wi-Fi association and the fixed settle delay took \SI{963}{\milli\second} and \SI{1000}{\milli\second}. Once the device had an address, the capsule exchange usually completed in \SI{18}{\milli\second}. One exchange took \SI{3000}{\milli\second}, and the benchmark did not identify the source of that delay. Server construction took a median \SI{137.5}{\micro\second}. Client processing required about \SI{36.0}{\milli\second}, including \SI{12.9}{\milli\second} for ML-DSA-44 verification, \SI{5.2}{\milli\second} for ML-KEM-512 decapsulation, and \SI{1.6}{\milli\second} for HKDF and pool injection.

The \SI{3252}{\byte} response requires three IP fragments at a \SI{1500}{\byte} IPv4 MTU. All 30 exchanges completed. This sample does not estimate loss under fragmentation. Memory remains tight in this post-quantum build: static data occupy \SI{184712}{\byte}, or 93.95\%, of Zephyr's \SI{192}{\kibi\byte} static DRAM region, alongside a peak measured heap of \SI{48216}{\byte}. The benchmark did not instrument energy.

\subsection{SRAM Startup State Capture}
\label{sec:puf_evaluation_sec}

Our SRAM startup data remains preliminary. A dedicated \texttt{.noinit} region lets the firmware copy \SI{4096}{\byte} at Zephyr's earliest portable initialization level. Before each RTS/EN reset, the firmware fills and verifies the region with \texttt{00}, \texttt{ff}, \texttt{aa}, or \texttt{55}. We interleaved the patterns and tested each 10 times on both boards.

On the primary ESP32, mean agreement with the preceding pattern ranged from 49.01\% to 50.99\%. Mean pairwise Hamming distance across the 40 captures was 5.98\%. On the ESP32-S3, the corresponding values were 49.40\% to 50.50\% and 5.10\%. All 40 hashes were distinct on each board. The captures did not follow the values written before reset, which excludes simple retention as the cause of their variation under RTS/EN.

The capture hook worked, and direct retention of the preceding pattern does not explain the observed variation. The experiment did not remove power, so it cannot establish cold startup entropy or device specificity. SRAM therefore enters the combiner only as uncredited auxiliary material ($P$). Crediting it would require controlled power removal across boards, off durations, and environments, followed by measurements of bias, stability, uniqueness, and min-entropy~\cite{holcomb2008power,herder2014physical}.

\section{Limitations}
\label{sec:limitations}

The radio measurements begin after UDP receipt, so the recorded arrival deltas combine radio behavior with interrupts, driver processing, network stack execution, and scheduling. This observation point prevents us from assigning conditional entropy to packet timing or separating analog effects from software variation. Channel state information (CSI), hardware timestamps, and temperature telemetry would provide better visibility. The WDEV rate also remains specific to the admitted ESP32 state and adversary model. Although the isolated \SI{1}{\mega\byte} stream passes Rabbit, unexplained flags across the full \SI{256}{\mega\byte} captures justify conditioning all raw WDEV output before consumption.

The hardware evidence covers one ESP32 with revision~1.0 silicon and selected checks on one ESP32-S3 with revision~0.2 silicon. Both boards remained powered through USB during RTS/EN resets, and the S3 did not repeat the full statistical battery or capsule benchmark. Establishing behavior across a population will require more boards and operating conditions, while a cold startup claim for SRAM will require controlled power removal over different off times, voltages, and temperatures.

The capsule exchange avoids random generation by the client. It depends on provisioned keys and persistent replay state. The client stores its decapsulation private key and the trusted node's verification key; the trusted node stores the corresponding client public key and selects it from the identity in \texttt{BOOT\_HELLO}. Only the client holding the matching private key can recover the capsule. The trusted node also knows its material. Because the prototype's boot counter is volatile, production firmware must retain accepted capsule sequence numbers in nonvolatile storage to reject replays across restarts.

\section{Conclusions}
\label{sec:conclusions}

Boot entropy should not depend on a single opaque source. We presented an architecture based on defense in depth that combines device startup state, an admitted local hardware RNG response, and a protected remote capsule while keeping entropy accounting separate from mixing. Under the assumptions in Section~\ref{sec:combiner}, one sufficiently unpredictable credited input is enough to keep the derived seed unpredictable given the other inputs and the public transcript.

The measurements show why admission is central to this design. ESP32 output remains statistically plausible even in the documented RF-disabled pseudorandom state, so firmware must establish the operating state before crediting WDEV. Once admitted under the RF-active model, fixed public bursts provide repeatable acquisition windows whose WDEV outputs varied within an association and across resets on both processors. On the primary ESP32, the output stream estimate extrapolates to approximately \SI{59}{\kilo\bit} of min-entropy per window before extraction into a 256-bit seed.

The capsule path complements this local root with material protected by a provisioned client key. All 30 measured exchanges reached the configured 128-bit accounting state, and association and DHCP, rather than client cryptography, governed their latency. The resulting design bootstraps an embedded random pool without accepting unchecked PRNG output as entropy.

\paragraph*{Code and Data Availability.}
The v0.1.1 release contains the evaluated firmware, scripts, and supporting measurement data.

\printbibliography

@inproceedings{heninger2012mining,
  title={Mining your Ps and Qs: Detection of widespread weak keys in network devices},
  author={Heninger, Nadia and Durumeric, Zakir and Wustrow, Eric and Halderman, J Alex},
  booktitle={21st USENIX Security Symposium (USENIX Security 12)},
  pages={205--220},
  year={2012}
}

@article{vassilev2016entropy,
  title={Entropy as a Service: Unlocking Cryptography's Full Potential},
  author={Vassilev, Apostol and Staples, Robert},
  journal={Computer},
  volume={49},
  number={9},
  pages={98--102},
  year={2016},
  publisher={IEEE}
}

@article{blanco2026post,
  title={Post-Quantum Entropy as a Service for Embedded Systems},
  author={Blanco-Romero, Javier and Garcia-Ni{\~n}o, Yuri Melissa and Mendoza, Florina Almenares and D{\'\i}az-S{\'a}nchez, Daniel and Garc{\'\i}a-Rubio, Carlos and Campo, Celeste},
  journal={Sensors},
  volume={26},
  number={9},
  pages={2737},
  year={2026},
  publisher={MDPI AG}
}

@techreport{nist2024fips203,
  title = {{Module-Lattice-Based Key-Encapsulation Mechanism Standard}},
  author = {{National Institute of Standards and Technology}},
  institution = {National Institute of Standards and Technology},
  type = {{Federal Information Processing Standards Publication}},
  number = {203},
  year = {2024},
  doi = {10.6028/NIST.FIPS.203},
  url = {https://doi.org/10.6028/NIST.FIPS.203}
}

@techreport{nist2024fips204,
  title = {{Module-Lattice-Based Digital Signature Standard}},
  author = {{National Institute of Standards and Technology}},
  institution = {National Institute of Standards and Technology},
  type = {{Federal Information Processing Standards Publication}},
  number = {204},
  year = {2024},
  doi = {10.6028/NIST.FIPS.204},
  url = {https://doi.org/10.6028/NIST.FIPS.204}
}

@techreport{sonmez2016recommendation,
  title={Recommendation for the Entropy Sources Used for Random Bit Generation},
  author={Turan, Meltem S{\"o}nmez and Barker, Elaine and Kelsey, John and McKay, Kerry A. and Baish, Mary L. and Boyle, Michael},
  institution={National Institute of Standards and Technology},
  type={NIST Special Publication},
  number={800-90B},
  year={2018},
  month={1},
  doi={10.6028/NIST.SP.800-90B},
  url={https://doi.org/10.6028/NIST.SP.800-90B}
}

@techreport{peter2024ais2031,
  title={{A Proposal for Functionality Classes for Random Number Generators}},
  author={Peter, Matthias and Schindler, Werner},
  institution={Bundesamt f{\"u}r Sicherheit in der Informationstechnik},
  type={{AIS 20 / AIS 31}},
  version={3.0},
  date={2024-09-10},
  url={https://www.bsi.bund.de/SharedDocs/Downloads/EN/BSI/Certification/Interpretations/AIS_31_Functionality_classes_for_random_number_generators_e_2024.pdf}
}

@inproceedings{saarinen2022sp,
  title={SP 800--22 and GM/T 0005--2012 tests: clearly obsolete, possibly harmful},
  author={Saarinen, Markku-Juhani O},
  booktitle={2022 IEEE European Symposium on Security and Privacy Workshops (EuroS\&PW)},
  pages={31--37},
  year={2022},
  organization={IEEE}
}

@inproceedings{dodis2013security,
  title={Security analysis of pseudo-random number generators with input: /dev/random is not robust},
  author={Dodis, Yevgeniy and Pointcheval, David and Ruhault, Sylvain and Vergniaud, Damien and Wichs, Daniel},
  booktitle={Proceedings of the 2013 ACM SIGSAC conference on Computer \& communications security},
  pages={647--658},
  year={2013}
}

@misc{rfc8446,
    series =    {Request for Comments},
    number =    8446,
    howpublished =  {RFC 8446},
    publisher = {RFC Editor},
    doi =       {10.17487/RFC8446},
    url =       {https://www.rfc-editor.org/info/rfc8446},
    author =    {Eric Rescorla},
    title =     {{The Transport Layer Security (TLS) Protocol Version 1.3}},
    pagetotal = 160,
    year =      2018,
    month =     aug,
}

@misc{rfc9147,
    series =    {Request for Comments},
    number =    9147,
    howpublished =  {RFC 9147},
    publisher = {RFC Editor},
    doi =       {10.17487/RFC9147},
    url =       {https://www.rfc-editor.org/info/rfc9147},
    author =    {Eric Rescorla and Hannes Tschofenig and Nagendra Modadugu},
    title =     {{The Datagram Transport Layer Security (DTLS) Protocol Version 1.3}},
    pagetotal = 61,
    year =      2022,
    month =     apr,
}

@misc{rfc9180,
    series =    {Request for Comments},
    number =    9180,
    howpublished =  {RFC 9180},
    publisher = {RFC Editor},
    doi =       {10.17487/RFC9180},
    url =       {https://www.rfc-editor.org/info/rfc9180},
    author =    {Richard Barnes and Karthikeyan Bhargavan and Benjamin Lipp and Christopher A. Wood},
    title =     {{Hybrid Public Key Encryption}},
    pagetotal = 107,
    year =      2022,
    month =     feb,
}

@misc{rfc5869,
    series =    {Request for Comments},
    number =    5869,
    howpublished =  {RFC 5869},
    publisher = {RFC Editor},
    doi =       {10.17487/RFC5869},
    url =       {https://www.rfc-editor.org/info/rfc5869},
    author =    {Hugo Krawczyk and Pasi Eronen},
    title =     {{HMAC-based Extract-and-Expand Key Derivation Function (HKDF)}},
    pagetotal = 14,
    year =      2010,
    month =     may,
}

@misc{rfc7748,
    series =    {Request for Comments},
    number =    7748,
    howpublished =  {RFC 7748},
    publisher = {RFC Editor},
    doi =       {10.17487/RFC7748},
    url =       {https://www.rfc-editor.org/info/rfc7748},
    author =    {Adam Langley and Mike Hamburg and Sean Turner},
    title =     {{Elliptic Curves for Security}},
    pagetotal = 22,
    year =      2016,
    month =     jan,
}

@misc{rfc8937,
    series =    {Request for Comments},
    number =    8937,
    howpublished =  {RFC 8937},
    publisher = {RFC Editor},
    doi =       {10.17487/RFC8937},
    url =       {https://www.rfc-editor.org/info/rfc8937},
    author =    {Cas Cremers and Luke Garratt and Stanislav V. Smyshlyaev and Nick Sullivan and Christopher A. Wood},
    title =     {{Randomness Improvements for Security Protocols}},
    pagetotal = 9,
    year =      2020,
    month =     oct,
}

@inproceedings{guajardo2007fpga,
  title={FPGA intrinsic PUFs and their use for IP protection},
  author={Guajardo, Jorge and Kumar, Sandeep S and Schrijen, Geert-Jan and Tuyls, Pim},
  booktitle={International workshop on cryptographic hardware and embedded systems},
  pages={63--80},
  year={2007},
  organization={Springer}
}

@inproceedings{markettos2009frequency,
  title={The frequency injection attack on ring-oscillator-based true random number generators},
  author={Markettos, A Theodore and Moore, Simon W},
  booktitle={International Workshop on Cryptographic Hardware and Embedded Systems},
  pages={317--331},
  year={2009},
  organization={Springer}
}

@inproceedings{dodis2004fuzzy,
  title={Fuzzy extractors: How to generate strong keys from biometrics and other noisy data},
  author={Dodis, Yevgeniy and Reyzin, Leonid and Smith, Adam},
  booktitle={International conference on the theory and applications of cryptographic techniques},
  pages={523--540},
  year={2004},
  organization={Springer}
}

@inproceedings{mathur2008radio,
  title={Radio-telepathy: extracting a secret key from an unauthenticated wireless channel},
  author={Mathur, Suhas and Trappe, Wade and Mandayam, Narayan and Ye, Chunxuan and Reznik, Alex},
  booktitle={Proceedings of the 14th ACM international conference on Mobile computing and networking},
  pages={128--139},
  year={2008}
}

@article{mukherjee2014principles,
  title={Principles of physical layer security in multiuser wireless networks: A survey},
  author={Mukherjee, Amitav and Fakoorian, S Ali A and Huang, Jing and Swindlehurst, A Lee},
  journal={IEEE Communications Surveys \& Tutorials},
  volume={16},
  number={3},
  pages={1550--1573},
  year={2014},
  publisher={IEEE}
}

@article{jiao2019physical,
  title={Physical Layer Key Generation in 5G Wireless Networks},
  author={Jiao, Long and Wang, Ning and Wang, Pu and Alipour-Fanid, Amir and Tang, Jie and Zeng, Kai},
  journal={arXiv preprint arXiv:1908.10362},
  year={2019},
  url={https://arxiv.org/abs/1908.10362}
}

@article{aldaghri2019physical,
  title={Physical Layer Secret Key Generation in Static Environments},
  author={Aldaghri, Nasser and Mahdavifar, Hessam},
  journal={arXiv preprint arXiv:1908.03637},
  year={2019},
  url={https://arxiv.org/abs/1908.03637}
}

@article{yang2022risinduced,
  title={Reconfigurable Intelligent Surface-Induced Randomness for mmWave Key Generation},
  author={Yang, Shubo and Han, Han and Liu, Yihong and Guo, Weisi and Pang, Zhibo and Zhang, Lei},
  journal={arXiv preprint arXiv:2111.00428},
  year={2022},
  url={https://arxiv.org/abs/2111.00428}
}

@article{li2022risconstructive,
  title={Reconfigurable Intelligent Surface for Physical Layer Key Generation: Constructive or Destructive?},
  author={Li, Guyue and Hu, Lei and Staat, Paul and Elders-Boll, Harald and Zenger, Christian and Paar, Christof and Hu, Aiqun},
  journal={arXiv preprint arXiv:2112.10043},
  year={2022},
  url={https://arxiv.org/abs/2112.10043}
}

@article{suzuki2018wirelesstrng,
  title={A True Random Number Generator Method Embedded in Wireless Communication Systems},
  author={Suzuki, Toshinori and Kaminaga, Masahiro},
  journal={arXiv preprint arXiv:1811.10783},
  year={2018},
  url={https://arxiv.org/abs/1811.10783}
}

@article{kietzmann2021guideline,
  title={A Guideline on Pseudorandom Number Generation (PRNG) in the IoT},
  author={Kietzmann, Peter and Schmidt, Thomas C. and W{\"a}hlisch, Matthias},
  journal={ACM Computing Surveys},
  volume={54},
  number={6},
  pages={1--38},
  year={2021},
  publisher={ACM},
  doi={10.1145/3453159}
}

@inproceedings{tillmanns2020firmware,
  title={Firmware Insider: Bluetooth Randomness is Mostly Random},
  author={Tillmanns, J{\"o}rn and Classen, Jiska and Rohrbach, Felix and Hollick, Matthias},
  booktitle={14th USENIX Workshop on Offensive Technologies (WOOT 20)},
  year={2020},
  publisher={USENIX Association}
}

@inproceedings{aumasson2013blake2,
  title={BLAKE2: simpler, smaller, fast as MD5},
  author={Aumasson, Jean-Philippe and Neves, Samuel and Wilcox-O'Hearn, Zooko and Winnerlein, Christian},
  booktitle={International Conference on Applied Cryptography and Network Security},
  pages={119--135},
  year={2013},
  organization={Springer}
}

@article{l2007testu01,
  title={TestU01: AC library for empirical testing of random number generators},
  author={L'ecuyer, Pierre and Simard, Richard},
  journal={ACM Transactions on Mathematical Software (TOMS)},
  volume={33},
  number={4},
  pages={1--40},
  year={2007},
  publisher={ACM New York, NY, USA}
}

@misc{walker2008ent,
  title={{ENT}: A Pseudorandom Number Sequence Test Program},
  author={Walker, John},
  year={2008},
  url={https://www.fourmilab.ch/random/}
}

@misc{dotyhumphrey2019practrand,
  title={{PractRand} Random Number Test Suite},
  author={Doty-Humphrey, Chris},
  year={2019},
  url={https://pracrand.sourceforge.net/}
}

@manual{espressif2026esp32trm,
  title = {ESP32 Technical Reference Manual},
  organization = {Espressif Systems},
  edition = {Version 5.7},
  year = {2026},
  url = {https://www.espressif.com/en/support/documents/technical-documents}
}

@inproceedings{yilek2009private,
  title={When private keys are public: Results from the 2008 Debian OpenSSL vulnerability},
  author={Yilek, Scott and Rescorla, Eric and Shacham, Hovav and Enright, Brandon and Savage, Stefan},
  booktitle={Proceedings of the 9th ACM SIGCOMM Conference on Internet Measurement},
  pages={15--27},
  year={2009}
}

@inproceedings{bernstein2013factoring,
  title={Factoring RSA keys from certified smart cards: Coppersmith in the wild},
  author={Bernstein, Daniel J and Chang, Yun-An and Cheng, Chen-Mou and Chou, Li-Ping and Heninger, Nadia and Lange, Tanja and Van Someren, Nicko},
  booktitle={International Conference on the Theory and Application of Cryptology and Information Security},
  pages={341--360},
  year={2013},
  organization={Springer}
}

@inproceedings{checkoway2014practical,
  title={On the practical exploitability of dual $\{$EC$\}$ in $\{$TLS$\}$ implementations},
  author={Checkoway, Stephen and Niederhagen, Ruben and Everspaugh, Adam and Green, Matthew and Lange, Tanja and Ristenpart, Thomas and Bernstein, Daniel J and Maskiewicz, Jake and Shacham, Hovav and Fredrikson, Matthew},
  booktitle={23rd USENIX security symposium (USENIX security 14)},
  pages={319--335},
  year={2014}
}

@inproceedings{checkoway2016systematic,
  title={A systematic analysis of the juniper dual EC incident},
  author={Checkoway, Stephen and Maskiewicz, Jacob and Garman, Christina and Fried, Joshua and Cohney, Shaanan and Green, Matthew and Heninger, Nadia and Weinmann, Ralf-Philipp and Rescorla, Eric and Shacham, Hovav},
  booktitle={Proceedings of the 2016 ACM SIGSAC Conference on Computer and Communications Security},
  pages={468--479},
  year={2016}
}

@inproceedings{mowery2013welcome,
  title={Welcome to the Entropics: Boot-time entropy in embedded devices},
  author={Mowery, Keaton and Wei, Michael and Kohlbrenner, David and Shacham, Hovav and Swanson, Steven},
  booktitle={2013 IEEE Symposium on Security and Privacy},
  pages={589--603},
  year={2013},
  organization={IEEE}
}

@article{holcomb2008power,
  title={Power-up SRAM state as an identifying fingerprint and source of true random numbers},
  author={Holcomb, Daniel E and Burleson, Wayne P and Fu, Kevin},
  journal={IEEE Transactions on Computers},
  volume={58},
  number={9},
  pages={1198--1210},
  year={2008},
  publisher={IEEE}
}

@article{herder2014physical,
  title={Physical unclonable functions and applications: A tutorial},
  author={Herder, Charles and Yu, Meng-Day and Koushanfar, Farinaz and Devadas, Srinivas},
  journal={Proceedings of the IEEE},
  volume={102},
  number={8},
  pages={1126--1141},
  year={2014},
  publisher={IEEE}
}

@inproceedings{barak2005model,
  title={A model and architecture for pseudo-random generation with applications to/dev/random},
  author={Barak, Boaz and Halevi, Shai},
  booktitle={Proceedings of the 12th ACM conference on Computer and communications security},
  pages={203--212},
  year={2005}
}

\end{document}